\documentclass[12pt]{article}
\usepackage{epsfig}
 
\newenvironment{Itemize}{\begin{list}{$\bullet$}%
{\setlength{\topsep}{0.2mm}\setlength{\partopsep}{0.2mm}%
\setlength{\itemsep}{0.2mm}\setlength{\parsep}{0.2mm}}}%
{\end{list}}
\newcounter{enumct}
\newenvironment{Enumerate}{\begin{list}{\arabic{enumct}.}%
{\usecounter{enumct}\setlength{\topsep}{0.2mm}%
\setlength{\partopsep}{0.2mm}\setlength{\itemsep}{0.2mm}%
\setlength{\parsep}{0.2mm}}}{\end{list}}
 
\newlength{\abstwidth}
\begin{document}
 
\sloppy
 
\pagestyle{empty}
 
\begin{flushright}
LU TP 96--21 \\
August 1996
\end{flushright}
 
\vspace{\fill}
 
\begin{center}
{\LARGE\bf Transverse Momentum as a Measure}\\[3mm]
{\LARGE\bf of Colour Topologies }\\[10mm]
{\Large E. Norrbin\footnote{emanuel@thep.lu.se} and %
T. Sj\"ostrand\footnote{torbjorn@thep.lu.se}} \\[3mm]
{\it Department of Theoretical Physics,}\\[1mm]
{\it Lund University, Lund, Sweden}
\end{center}
 
\vspace{\fill}
 
\begin{center}
{\bf Abstract}\\[2ex]
\begin{minipage}{\abstwidth}
Several distinct colour flow topologies are possible in 
multiparton configurations.  A method is proposed to find the 
correct topology, based on a minimization of the total 
transverse momentum of produced particles. This method is 
studied for three-jet $Z^0 \to q\overline{q} g$ and four-jet 
$W^+W^- \to q_1\overline{q}_2q_3\overline{q}_4$ events. It is 
shown how the basic picture is smeared, especially by parton-shower 
activity. The method therefore may not be sufficient on its 
own, but could still be a useful complement to others, and
e.g. help provide some handle on colour rearrangement 
effects.
\end{minipage}
\end{center}
 
\vspace{\fill}
 
\clearpage
\pagestyle{plain}
\setcounter{page}{1}

When high-energy processes produce multiparton sta\-tes,
it is generally believed that the confinement property of QCD
leads to the formation of colour flux tubes or vortex lines
spanned between the partons. These tubes/vortices are here called
strings, in anticipation of our use of hadronization based
on the string model \cite{Lundmod}. A quark or antiquark
is attached to one end of a string. A gluon is attached 
to two string pieces, one for its colour and one for its 
anticolour index, and thus corresponds to a kink on the string. 

The simplest kind of events, 
$e^+e^- \to \gamma^*/Z^0 \to q\overline{q}$, gives a single string
stretched between the $q$ and the $\overline{q}$. The direction of the 
colour flow can, in principle, be distinguished by flavour correlations 
\cite{BA+HUB}, but will not be studied here. In next order, 
$q\overline{q} g$ events correspond to a string stretched from the $q$ 
via the $g$ to the $\overline{q}$. The colour topology is unique, but
experimentally it is not normally known which of the three jets is
the gluon one, so this gives a threefold experimental ambiguity.  

From four partons onwards, true ambiguities of the topology 
exist, even when the identity of the partons is known. In 
$q\overline{q} gg$ events, the string can be drawn from the q to either 
of the two gluons, on to the other gluon and then to the $\overline{q}$.
This gives two possible topologies. A third topology, not expected
to leading order in $N_C$, is when one string runs directly 
between the $q$ and $\overline{q}$ and another string in a closed loop
between the two gluons. There is an experimental ambiguity, in
picking the two quarks among the four partons, which gives a further
factor of six, i.e. a total of eighteen possible topologies
(reduced to fifteen if single and double strings are not 
distinguished).

Another four-jet final state is obtained in the process
$e^+e^- \to W^+W^- \to q_1\overline{q}_2q_3\overline{q}_4$. 
Since the $W$'s are colour singlets, in principle each of 
$q_1\overline{q}_2$ and $q_3\overline{q}_4$ form a separate colour 
singlet string. However, by soft gluon exchange or some other 
mechanism, alternatively $q_1\overline{q}_4$ and $q_3\overline{q}_2$ 
may form two singlets. Since it would not normally be known which of 
the four jets are quarks and which antiquarks, there is a total of 
three experimental pairings of four jets, where the third corresponds 
to the unphysical flavour sets $q_1q_3$ and 
$\overline{q}_2\overline{q}_4$.

Among the topologies above, only the three-jet $q\overline{q}g$ events 
have been studied in detail. It has been shown that the string 
approach here correctly predicts the topology of particle flow,
with a dip in the angular region between the $q$ and $\overline{q}$
jets \cite{stringtheory,stringdata}. This comes about because the 
$qg$ and $\overline{q}g$ string pieces produce (soft) particles in 
the respective angular ranges, while there is no string directly
between the $q$ and $\overline{q}$. The same effect is also obtained as 
a consequence of colour coherence in perturbative soft-gluon emission 
\cite{dipoleeffect} --- normally these two approaches give the same 
qualitative picture. Although by now LEP~1 has produced large samples 
of four-jet events, the energy flow between these jets have not been 
studied in detail, presumably because of the large number of possible 
topologies.

LEP~2 will provide four-jets from $W$ pairs, and here the issue of 
colour topology may become of great importance \cite{GPZ,SjoValery}.
If, by colour rearrangement, an original $q_1\overline{q}_2$ plus 
$q_3\overline{q}_4$ colour singlet configuration is turned into a
$q_1\overline{q}_4$ plus $q_3\overline{q}_2$ one, particle production 
will be somewhat different. Methods to determine the $W$ mass from 
LEP~2 four-jet events will then give different results. There is 
more than one model of the colour rearrangement process; therefore the
uncertainty on the $W$ mass could be as high as 100~MeV 
\cite{workshop}, i.e. larger than the expected statistical error  
of the order of 40~MeV. The final number will then be dominated by the
mixed hadronic--leptonic channel $W^+W^- \to \ell\nu_{\ell}q\overline{q}$,  
which has about the same statistics. The hadronic events could be 
recuperated if colour rearrangement effects could be diagnosed from 
the data itself. Additionally, the ability to distinguish between
various reconnection scenarios could provide information on the 
nature of the QCD vacuum, and therefore be of fundamental interest.
Unfortunately, realistic reconnection scenarios give only very minute
effects on the data (remember that the effect on the $W$ mass is 
at most of the order of one per mille), and so attempts to find useful 
signals have been rather unsuccessful \cite{SjoValery}. There is 
only one claim for having found a signal \cite{GJari} where, 
in a few events, a central rapidity gap separates the particle 
production from two low-mass colour singlet reconnected systems. 
It has turned out, however, that neglected angular correlations in 
the W pair decays tend to reduce this signal to the border of 
observability \cite{GPW,JGerrata}. Therefore one would like to devise 
alternative methods to diagnose the appearance of colour 
rearrangement. 
 
We now want to propose and study a method to select the correct string 
topology. The starting point is the observation that, were it not
for various smearing effects, hadrons would be perfectly lined up 
with the string pieces spanned between the partons. In momentum
space the hadrons would appear along hyperbolae with the respective 
endpoint parton directions as asymptotes. In a frame where a string 
piece has no transverse motion, i.e. where the two endpoint partons 
are moving apart back-to-back, the hadrons produced by this piece would 
have vanishing transverse momentum. A hadron could then successively be 
boosted to the rest frame of any parton pair until the pair is found for 
which the particle $p_{\perp}$ is vanishing. When smearing is introduced, 
this $p_{\perp}$ will no longer be vanishing, but it still would be 
reasonable that a ``best bet'' is to assign each hadron to the string 
piece with respect to which it has smallest $p_{\perp}$. The most likely 
string configuration would be the one where the sum of all hadron 
$p_{\perp}$'s is minimal. While the fluctuations for each single hadron 
is too large for usefulness, it could be hoped that the net effect of 
all the hadrons would be to single out the correct topology. 

To be more precise, here is the scheme proposed, to be carried out 
for each event:
\begin{Enumerate}
\item Use some jet clustering algorithm to identify the directions
of the number of jets that should be used for the current application.
\item Enumerate the possible colour flow topologies allowed
in the process. 
\item Form one $\sum p_{\perp}$ measure for each topology, where the
sum runs over all final-state particles in the event. For each particle, 
its $p_{\perp}$ is defined as the minimum of the $p_{\perp}$'s obtained 
by boosting the particle to the rest frame of each of the string pieces 
making up the current topology.
\item Identify the correct topology as the one with smallest 
$\sum p_{\perp}$.
\end{Enumerate}

Several effects could smear the simple picture and lead to incorrect
conclusions. The main ones are:
\begin{Itemize}
\item Errors in the reconstruction of jet directions.
\item Additional $p_{\perp}$ caused by perturbative QCD branchings,
predominantly gluon emission.
\item The $p_{\perp}$ generated by the fragmentation process.
\item Secondary decays of unstable hadrons.
\end{Itemize} 

A convenient test bed for the relative importance of these effects 
is provided by $q\overline{q}g$ events. In symmetric three-jet events
at $Z^0$ energies, without any parton-shower activity, the 
gluon is correctly identified in 87\% of the cases. This should be
compared with the 33\% expected from a random picking among the
three alternatives. When complete $Z^0$ events events are
generated (with {\sc Pythia}~5.7 and {\sc Jetset}~7.4 \cite{Manual}),
three jets are found (in 10--20\% of the events) and the method
above is used, the success rate is around 55--60\%. The ``correct''
answer is here found by tracing the original quark and antiquark
through the shower history and associating them with the jets they 
are closest to in angle. The conclusion of this kind of studies is 
that the method does work, though maybe not as well as one might 
have hoped, and that the main cause of errors is the perturbative 
gluon emission.

The studies can be extended to four-jet $q\overline{q}gg$ events,
although the large number of colour flow topologies here makes the
method rather inefficient for selecting the correct topology. A more
modest objective is to establish differences between string and  
independent fragmentation \cite{indep} models. In the latter
approach, particle production is aligned along the direction 
connecting a parton with the origin in the c.m. frame of the event,
again with some smearing effects. This approach is already
disfavoured by the three-jet studies \cite{stringdata}, but
is a convenient reference. We have not completed a full
study, but can illustrate with results for a simple four-jet
cross topology with $q$ and $\overline{q}$ back-to-back. 
Fig.~\ref{fourjets}a gives the difference in $\sum p_{\perp}$
between the worst topology (where none of the string pieces
run the way assumed) and the right one. Results are for
charged particles, and the $\sum p_{\perp}$ has been normalized to the 
number $N$ of charged particles per event. The independent
fragmentation distribution is symmetric around the origin,
as it should,
while the string approach shows a clear offset. Without knowledge 
of the correct answer, one is reduced to studying a measure such
as the difference between the maximal and the minimal $\sum p_{\perp}$
among all the twelve possible topologies, Fig.~\ref{fourjets}b. 
By construction this is a positive number, also for independent 
fragmentation, but an additional shift is visible for the string 
approach. Identification of one or both quarks, e.g. by 
$b$ quark tagging or energy ordering (quarks are likely to have
higher energy than gluons), would cut down on the number of 
topologies to be compared and therefore enhance the signal.
Some experimental studies along these lines could therefore be 
interesting.

We now turn to the main application of this letter, namely 
four-jet $W^+W^-$ events. Since the two $W$ decay vertices
are less than or of the order of 0.1~fm apart, while typical
hadronic distances are of the order of 1~fm, the two $W$
decays occur almost on top of each other. QCD 
interconnection effects could appear at all stages of the process,
namely in the original perturbative parton cascades, in the
subsequent soft hadronization stage, and in the final hadronic
state. It can be shown that perturbative effects are strongly
suppressed \cite{SjoValery}, but no similar arguments hold for the
other two. Bose-Einstein effects in the hadronic state could 
be the largest individual source of $W$ mass uncertainty
\cite{LeifTor}, but it is the least well studied. The presence of
such Bose-Einstein effects presumably could be established from the
data itself, while reconnection in the hadronization stage 
is less easy to diagnose. We will study whether the $\sum p_{\perp}$ 
measure offers any help here.
 
The reconnection models used as references in this work are:
\begin{Enumerate}
\item Reconnection after the perturbative shower stage but before 
the hadronization, with reconnection occuring at the `origin' of the 
showering systems. This `intermediate' model is the simplest of the 
more realistic ones.
\item Reconnection when strings overlap based on cylindrical geometry. 
A `bag model' based on a type I superconductor analogy. The 
reconnection probability is proportional to the overlap integral 
between the field strengths, with each field having a Gaussian 
fall-off in the transverse direction, the radius being about 0.5~fm. 
The model contains a free strength parameter that can be modified to 
give any reconnection probability.
\item Reconnection when strings cross. In this model the strings mimic 
the behaviour of the vortex lines in a type II superconductor, where 
all topological information is given by a one-dimensional region at 
the core of the string.
\item Reconnections occur in such a way that the string `length' is 
minimized. As a measure of this length the so-called $\lambda$-measure 
is used, which essentially represents the rapidity range for particle 
production counted along the string. This can also be seen as a measure 
of the potential energy of the string.
\end{Enumerate}
Models 1 through 3 is described in \cite{SjoValery} and the last in 
\cite{GJari}. Further models have been proposed \cite{morerecon}.

For the study, events should have a clear four-jet structure. To achieve 
this we demand that each jet must have some minimum energy and that the 
angle between any jet pair must not be too small \cite{SjoValery}. 
When applied to the expected statistics of LEP~2, the number of events 
left after the cuts will be about 2500 per experiment; therefore 
statistics will be a problem when different models are compared with 
each other.

Three different algorithms are used to identify which jet pairs belong
together:
\begin{Enumerate}
\item The $q_1\overline{q}_2q_3\overline{q}_4$ configuration before 
parton showers can be matched one-to-one with the reconstructed jets 
after hadronization. This is done by minimizing the products of the 
four (jet+$q$) invariant masses. The original quark information is not 
available in an experimental situation, so this measure can only be 
used as a theory reference.
\item The invariant mass of jet pairs from the same $W$ should be close to 
the known $W$-mass of about 80~GeV. Among the three possible jet pairings,
therefore the one is selected which has minimal
$|m_{ij}-80|+|m_{kl}-80|$, where $i,j,k,l$ are the four jets.
We have picked this method rather than a few similar ones since it has
(marginally) the best correlation with the reference method above. 
This method is mainly probing the electroweak aspect of $W$ pair
production, namely the $W$ mass spectrum, while it should be less 
dependent on the QCD stages of showering and hadronization. 
\item The $\sum p_{\perp}$ method introduced in this letter provides an 
alternative measure, that rather should be sensitive to the QCD
stages and less so to the electroweak one. Without colour reconnection
it should (hopefully) agree with the previous two, while it could
give interesting differences if reconnection occurs.
\end{Enumerate}

The agreement between these three methods is shown in Table~\ref{agree},
with and without reconnection, the former for model 1. 
As should be expected, algorithm 2 comes
close to the ``correct'' answer of number 1, and is not significantly
affected by colour rearrangement. The $\sum p_{\perp}$ method, 
algorithm 3, shows the expected dependence on colour rearrangement, 
with a smaller success rate when colour rearrangement is included. 
Note, however, that the success rate does not drop below the naive 
33\% number, indicating that the $\sum p_{\perp}$ method is also 
picking up other aspects of events, such as the jet topology. 
The results in the table are for a 170~GeV energy, but we do not 
expect a significant energy dependence.

Both methods 2 and 3 can be applied to data, so therefore the 
correlation is an observable. The rate of reconnection 
could be extracted, by interpolation between the two extremes of no and
complete colour rearrangement. Statistically it should be feasible to
establish a signal for reconnections, if they occur at a rate above
the 10--20\% level. The systematic errors on the correlation method may 
be large, however, especially for the model-dependent change when 
reconnection is included. It is therefore important to study whether 
a differential distribution would better highlight qualitative 
differences. 

Algorithm 2 can be used to identify the best hypothesis for which
jets should be paired to form the two $W$'s, and also the worst 
hypothesis, where $|m_{ij}-80|+|m_{kl}-80|$ is maximal. The 
$\sum p_{\perp}$ can be calculated for each of these two extremes.
When the strings are reconnected, the first sum should
increase and the second one decrease relative to the 
no-reconnection case. The signal is therefore
enhanced by making use of the difference, $\Delta = %
(\sum p_{\perp})_{\mathrm{worst}} - (\sum p_{\perp})_{\mathrm{best}}$,
which should decrease in case of reconnection. The subtraction 
furthermore has the advantage of removing some spurious 
fluctuations from the comparison. The main example is high-momentum 
particles, where the assumed string hyperbolae  attach well
with the four jet directions and therefore all three string hypotheses
give the same contribution to $\sum p_{\perp}$. (Our studies show that 
particles with momenta above 3~GeV add little to the discrimination
between the string hypotheses.)  

The $\Delta$ measure is plotted for models 1--4 in Fig.~\ref{delta}. 
Note that the results for models 1 and 4 correspond to 100\%
reconnection, while the reconnection rate in models 2 and 3 is 
about 30\%. (The reconnection fraction could be varied in all 
models, but the effects of reconnection are not linear in this
fraction for models 2 and 3, so a reasonable value is preferred 
here.) All reconnection models show the expected shift
towards smaller $\Delta$ values, and the magnitude of the shift
is comparable once differences in assumed reconnection fractions
are removed. Remaining differences imply that one cannot
model-independently extract a reconnection rate from the data.  
The signal for reconnection may be enhanced, compared with the results 
of Table~\ref{agree}, by cuts on $\Delta$, e.g. by only considering
the fraction of events with $\Delta < 0$. This is at the price of a
reduced statistics, however, so the balance is not so clear. 
It may be better to use measures that gauge the full shape of the 
curve, given that the physics of the no-reconnection scenario is
presumed well-known (by extrapolation from the $Z^0$ results).

Prospects look promising to diagnose colour rearrangement along 
these lines but, as before, the combination of low effects and small
statistics could give marginal results. Furthermore, hopes should 
not be raised too high that this would immediately imply
a scheme to correct a $W$ mass measurements for the reconnection
effects: of the models above, number 3 shifts the $W$ mass 
downwards while the others shift it upwards \cite{SjoValery,workshop}, 
and yet they all shift the $\Delta$ distribution in the same 
direction. Clearly the study of reconnection effects ultimately 
must be based on a host of different measures, the 
$\sum p_{\perp}$ one and others.

It may be of some interest to understand why effects are not larger.
Several simulations have been performed with various simplified
toy models to study this issue \cite{Emanuel}. It turns out that 
there are two main mechanisms that smear distributions and make 
them less easily distinguished. One is parton showers, just as 
for the $Z^0 \to q \overline{q} g$ process studied above. 
The other is the geometry of the process, namely that the helicity
structure of the $W^+W^- \to q_1 \overline{q}_2 q_3 \overline{q}_4$
process is such that $q_1$ and $q_3$ tend to go in the same 
general direction, as do $\overline{q}_2$ and $\overline{q}_4$
\cite{GPW}. The overall change of string topology 
by reconnection therefore is not as drastic
as if the $q_1$ had tended to be close to $\overline{q}_4$
and vice versa. Had it been possible to remove these effects, 
i.e. study events without shower activity and with $q_1$ and
$\overline{q}_4$ reasonably close in angle, the original and the 
colour-reconnected $\Delta$ distributions would almost completely
separate. In practice, only a modest reduction of shower activity 
could be obtained by requiring that all four jets be reasonably 
narrow, while tagging of quark vs. antiquark (e.g. with charm) 
would leave very few events. Therefore no simple solutions
have been found.

In summary, we have introduced a $\sum p_{\perp}$ measure as a 
diagnostic of the colour topology of hadronic events. The 
fuzzy nature of hadronic final states somewhat limits the
usefulness of the method. In particular, the more drastic
effects associated with perturbative gluon emission tend to
obscure the subtler effects of different colour topologies. Therefore
the $\sum p_{\perp}$ measure is no panacea, but could still be
a useful addition to the (not so large) tool box of methods
to characterize the nonperturbative stage of hadronic events.
Applications include three- and four-jet events at $Z^0$ energies 
and, in particular, $W$ pair decay to four jets at LEP~2. In the
latter process, it could be possible to detect the effects of
colour reconnection with this approach. Further details on the
studies reported here may be found in \cite{Emanuel}.

\begin{table}
\begin{center}
\begin{tabular}{|c|cc||c|cc|}
\hline
\multicolumn{3}{|c||}{no reconnection} &
\multicolumn{3}{c|}{with reconnection (model 1)} \\ \hline
algorithm   & 1    & 2    & algorithm   & 1    & 2     \\ \hline
2  & 85\% & --   & 2  & 85\% & --    \\
3  & 68\% & 65\% & 3  & 46\% & 46\%  \\ \hline
\end{tabular}\vspace{2mm}
\caption[]{Fraction of agreement in jet pair identification 
between the three algorithms.}
\label{agree}
\end{center}
\end{table}

\begin{figure}
\begin{center}
\mbox{\epsfig{file=fig1a.eps, width=70mm}\hspace{10mm}%
\epsfig{file=fig1b.eps, width=70mm}}\\
\mbox{(a)\hspace{70mm}(b)}
\end{center}
\caption[]{The distribution of $\sum p_{\perp}/N$ for \textbf{a}: 
worst$-$right, and \textbf{b}: maximal$-$minimal, 
for Lund string (full) and independent (dashed) fragmentation 
respectively.}
\label{fourjets}
\end{figure}

\begin{figure}
\begin{center}
\mbox{\epsfig{file=fig2a.eps, width=70mm}\hspace{10mm}%
\epsfig{file=fig2b.eps, width=70mm}}\\
\mbox{(a)\hspace{70mm}(b)}\vspace{10mm}\\
\mbox{\epsfig{file=fig2c.eps, width=70mm}\hspace{10mm}%
\epsfig{file=fig2d.eps, width=70mm}}\\
\mbox{(c)\hspace{70mm}(d)}
\end{center}
\caption[]{Distribution of $\Delta = %
(\sum p_{\perp})_{\mathrm{worst}} - (\sum p_{\perp})_{\mathrm{best}}$. 
Dashed lines are reconnected events according to models  
\textbf{a}: 1 (intermediate), \textbf{b}: 2 (bag model), 
\textbf{c}: 3 (type II superconductor), and \textbf{d}: 4 
($\lambda$-measure), while full lines always are without 
reconnection.}
\label{delta}
\end{figure}

\end{document}